\documentclass[%
 reprint,
 amsmath,amssymb,
 aps,
]{revtex4-2}

\usepackage{float}
\usepackage{placeins}

\usepackage{xcolor}
\definecolor{myteal}{HTML}{008080}

\usepackage[colorlinks=true,
            linkcolor=blue,
            citecolor=blue,
            urlcolor=blue]{hyperref}

\usepackage[colorlinks=true]{hyperref}

\usepackage{graphicx}% Include figure files
\usepackage{dcolumn}% Align table columns on decimal point
\usepackage{bm}% bold math
\begin{document}

\preprint{APS/123-QED}

%\title{Manuscript Title:\\with Forced Linebreak}% Force line breaks with \\

% \title{Laser-Induced Nanopore-Mediated Fast Crystallization of Dense Fused Silica by Molecular Dynamics coupled to Finite-Difference Time-Domain}

\title{From pore collapse to crystal growth: ultrafast laser-induced stishovite formation in nanoporous silica}

% ------- AUTHORS (no “and”, manual superscripts) -------

\author{%
  Aram Yedigaryan$^{1}$,
  Mohamed Yaseen Noor$^{2}$,
  Elena Kachan$^{1}$,
  Gabriel Calderon$^{2}$,
  Jinwoo Hwang$^{2}$,
  Enam Chowdhury$^{2}$,
  Jean\mbox{-}Philippe Colombier$^{1}$%
}

\affiliation{\vspace{1em}$^{1}$ Universit\'e Jean Monnet, CNRS, Institut d’Optique Graduate School, 
Laboratoire Hubert Curien UMR 5516, Saint-Etienne F-42023, France}

\affiliation{\vspace{0.3em}$^{2}$ Department of Materials Science and Engineering, The Ohio State University,
140W 19th Avenue, Columbus, 43210, Ohio, United States}

\date{\vspace{1em}\today}

\begin{abstract}
% The crystallization of amorphous solids under ultrafast laser irradiation represents a paradigm of non-equilibrium phase transitions, where the interplay between electromagnetic energy localization and atomic-scale dynamics remains largely uncharted.
% By using a multiscale framework that couples finite-difference time-domain simulations of nonlinear femtosecond laser pulse propagation with molecular dynamics of the atomic response, we demonstrate that field enhancement around the pores of nanoporous amorphous silica confines laser energy and drives a rapid pore collapse.
% In the silica structure containing a 4-nm pore, corresponding to a 7\% porosity, the enhanced local electromagnetic field led to a final equilibrium temperature 16\% higher than for the 2-nm pore (1\% porosity) due to bigger surface area, and 20\% higher than for the homogeneous medium. The heterogeneous energy localization in the 4-nm porous system served as a preferential nucleation site within the dense glass network leading to the ultrafast formation of stishovite crystal on a sub-nanosecond timescale, faster than in homogeneous silica.
% Such accelerated crystallization allows the phase transition to outpace pressure relaxation, which would otherwise inhibit stishovite formation under identical thermal loading. These theoretical results align well with experimental observations of femtosecond-laser-induced crystallization in confined geometries, and they show that electromagnetic hotspots in a nanoporous glass structure can be tailored to control solid-state transformations.
The crystallization of amorphous solids under ultrafast laser irradiation represents a paradigm of non-equilibrium phase transitions, where the interplay between electromagnetic energy localization and atomic-scale dynamics remains largely uncharted. By using a multiscale framework that couples finite-difference time-domain simulations of nonlinear femtosecond laser pulse propagation with molecular dynamics of the atomic response, we demonstrate that field enhancement around the pores of nanoporous amorphous silica confines laser energy and drives rapid pore collapse. In nanoporous silica, the enhanced local electromagnetic field leads to stronger energy absorption compared with smaller-pore and homogeneous systems. This heterogeneous energy localization provides preferential nucleation sites within the dense glass network, leading to ultrafast formation of stishovite on a sub-nanosecond timescale, faster than in homogeneous silica. This accelerated crystallization can outpace pressure relaxation making the transition to a high-pressure phase possible. These results are confirmed by experimental observations of femtosecond-laser-induced crystallization in confined geometries, and show that electromagnetic hotspots in nanoporous glass structures can be tailored to control solid-state transformations.
\end{abstract}

                             
\maketitle

%\tableofcontents

\section{Introduction}
% :\protect\\ The line
% break was forced \lowercase{via} \textbackslash\textbackslash}

Recreating the extreme conditions of Earth’s deep interior experimentally is a long-standing challenge in high-pressure physics. Dense silica polymorphs such as stishovite normally form only at pressures of tens of gigapascals and temperatures characteristic of the lower mantle \cite{PabstGregorova2013ElasticReview, stishov, miyahara, rehman2021phase}. Yet femtosecond lasers can deposit energy so rapidly that comparable thermodynamic states may be reached transiently within a solid. Silica (SiO$_2$) is uniquely suited to explore this possibility: in its amorphous form it consists of a continuous random network capable of densification and structural reorganization under ultrafast nonlinear excitation \cite{zheng2006densification, mishchik2012ultrafast, vignes2013thermomechanical}. The central question is therefore how such extreme states can nucleate and preserve a rare high-pressure crystalline phase shortly after ultrafast laser irradiation before the system relaxes back to disorder.

Femtosecond lasers deposit energy with nanometric confinement at the femtosecond timescale, driving matter far from equilibrium \cite{StoianColombier}. In silica, this leads to a rich variety of responses from smooth refractive-index modification (Type I) to self-organized nanogratings (Type II) and void formation (Type III), with more recent regimes such as Type X and Type C associated with porous and nanocrystalline phases \cite{shchedrina2025ultrafast, sakakura2020, Miura2000SpaceSelective}. These results reveal how strongly dielectrics respond to ultrafast excitation and raise the question of under what conditions a transiently heated glass can reorganize into a high-pressure crystal.

Although rare, laser-induced crystallization in silica has been observed under extreme conditions. Time-resolved X-ray diffraction experiments of laser-induced shock compression in silica revealed transient stishovite formation above 18 GPa on nanosecond timescales \cite{gleason2015ultrafast}, while molecular dynamics (MD) simulations showed pressure-driven transitions from tetrahedral to octahedral silicon coordination beyond 20 GPa \cite{trang}. These results indicate that crystallization is not anomalous, but rather emerges when the local energy density and pressure are sufficiently high. Glass-to-crystal transitions in silica can also occur due to laser-excitation-induced changes in the electrochemical potential \cite{poumellec2023electrostatic}.
All these conditions are especially relevant in nanoporous systems, where electromagnetic energy can be strongly confined around the pores.

Nanoporous structures formed during femtosecond irradiation are now recognized both as a distinct modification pathway and as a source of emergent functionality. Periodic nanopore arrays introduce birefringence and enable photonic applications \cite{lancry2013nanoporous, sakakura2020, shayeganrad2022high}. Their formation is driven by nonlinear photoionization and plasma-mediated nanocavitation \cite{rudenko2017role, lancry2011nanogratings}. Once created, these pores further reshape light-matter interaction: enhanced local-fields and heterogeneous stress relaxation around the pores amplify absorption and can modify crystallization kinetics \cite{lancry2016nanoscale, champion2013stress, Shimotsuma2003PRL}. However, the effect of nanopores on extreme excitation and phase transitions remains largely unexplored.

Addressing this question requires a multiscale framework that connects electromagnetic energy confinement to atomic-scale structural response. MD provides atomistic insight into nonequilibrium structural evolution \cite{iabbaden, iabbaden2, iabbaden2025nonequilibrium, ultrafastscience_nguyen, yedigaryan2025thermodynamic, dominic2024ultrafast, prudent2022high}, while the finite-difference time-domain (FDTD) method captures nonlinear light propagation and field enhancement in complex geometries and interfaces \cite{taflove2005, oskooi2010, rudenko2017spontaneous}. Combined, these approaches provide a unified description of electronic excitation, energy transfer, and lattice reorganization in laser-irradiated dielectrics.

Here, we develop a coupled FDTD-MD framework to model laser-induced crystallization in nanoporous silica. By explicitly incorporating excitation-induced bandgap narrowing into the photoionization rate calculation \cite{tsaturyan2024unraveling}, we demonstrate that electromagnetic confinement around the nanopore generates localized hotspots that drive rapid pore collapse and trigger faster crystallization into high-pressure stishovite on sub-nanosecond timescales.

This multiscale picture reveals that nanoscale porosity is not just a byproduct of laser modification but an important control parameter in ultrafast phase transitions, bridging nonlinear optics, shock physics, and crystallization theory in transparent dielectrics.

\section{Model}

% The major convenient mathematical model widely used in the computational physics community to manage laser-matter interaction at an atomistic scale is the Two-Temperature-Model (TTM) introduced by Anisimov \textit{et al.} \cite{anisimov1974electron}. In reality, the theoretical formulation of this model is based on coupling the diffusion equations for electrons and the lattice with the electron-phonon coupling term. Although the TTM takes into account the electronic effects, it is unable to describe thermoelastic stress and lattice density on the atomic level. Therefore, a hybrid model combining these two methods was preferred. Coupling of the atomistic lattice subsystem and the continuum electron subsystem was introduced in many previous works \cite{PhysRevB.44.567, PhysRevLett.71.1023, schaefer2002metal} and a further modified atomistic-continuum model was described by Ivanov and Zhigilei~\cite{ivanov2003combined}.
The modeling of target response to femtosecond laser irradiation is divided into two parts, the atomistic part which describes the lattice subsystem by MD, and the continuum part which accounts for the processes in the electron subsystem by Two-Temperature-Model (TTM).

Recently, a ``Femto3D" package was developed \cite{yuan2021ablation} based on TTM module in Large-scale Atomic/Molecular Massively Parallel Simulator (LAMMPS) \cite{thompson2022lammps} widely used for MD simulations. It is designed to model laser ablation in metallic targets and does not take into account complex 3D geometries and nonlinear processes like multiphoton and avalanche ionization in dielectrics. We adapt this package into a more versatile ``Femto3D+" package which includes the laser source term externally and can be used for metals, semiconductors and dielectrics. The laser source term can be calculated by FDTD, which is the case of our study.

\subsection{Two-Temperature Model coupled to Molecular Dynamics}

The general equations employed for our model are as follows:
\begin{equation}
\label{eq:conduction_El_El_NEW}
C_\text{e}(T_\text{e}) \frac{\partial T_\text{e}}{\partial t} = \nabla\left( K_\text{e}(T_\text{e}) \nabla T_\text{e} \right) \,-\, G_{\text{e-ph}}(T_\text{e})(T_\text{e} - T_\text{a}) \,+\, \mathcal{P}(t, \textbf{r})
% \tag{2.1}
\end{equation}
\begin{equation}
\label{eq:conduction_El_Lattice}
m_j \frac{d \mathbf{v}_j}{dt} = -\nabla_j U(\mathbf{r}_1, \ldots, \mathbf{r}_n) + \mathbf{F}^{\text{lang}}_j (T_\text{e} - T_\text{a})
% \tag{2.2}
\end{equation}
In these equations, $T_\text{e}$ and $T_\text{a}$ are the electron and lattice temperatures, $K_\text{e}$ and $C_\text{e}$ are the thermal conductivity and thermal capacity of the electron subsystem, $G_{\text{e-ph}}$ is the electron-phonon coupling factor, $\mathcal{P}$ is the absorbed laser power density distribution. In MD, $m_j$ is the mass of the atom $j$, $\mathbf{v}_j$ is its velocity, and $\mathbf{F}^{\text{lang}}_j$ is the Langevin force reflecting the energy transfer from the electronic subsystem. The internal force, $-\nabla_j U$, is the force acting on atom $j$ due to the interatomic potential $U$, where $\mathbf{r}_1$,...,$\mathbf{r}_n$ are the position vectors of all the atoms in the system.

\begin{figure}[!t]
\centering
\includegraphics[scale=0.342]{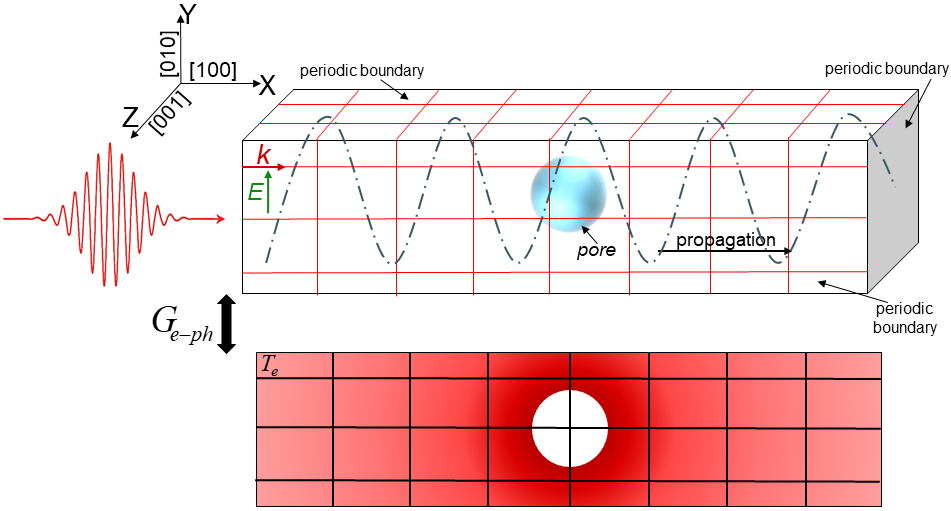}
\caption{Schematic representation of the coupled atomistic-continuum framework used in this work. The upper part depicts the lattice (atomic) subsystem, while the lower part represents the electronic subsystem responsible for laser-induced excitation and free-carrier dynamics.}
\label{Femto3D_scheme}
\end{figure}

Figure \ref{Femto3D_scheme} shows the scheme of the model. The periodic boundary conditions are set in all three directions. Thus, the atoms leaving from one boundary reenter the simulation box from the opposite boundary. This creates an effect of shock waves coming from other pores and replicates a laser-irradiated bulk medium under plane-wave irradiation.

At the atomic level, the electronic subsystem is treated as a continuous medium that effectively acts as a thermal reservoir in the MD simulation. Consequently, a canonical ensemble is appropriate for describing the lattice dynamics. Since the energy exchange between the electronic and lattice subsystems is governed by the coupling factor and their instantaneous temperatures, the Langevin thermostat is a suitable approach for modeling this interaction \cite{yuan2021ablation}.
% In the atomistic domain the electron subsystem is still considered as a continuum body, so for all the atoms in MD this continuum body can be considered as a heat bath. As a result, canonical ensemble is suitable in this case. Because in the model the temperature of the electron subsystem is changing every timestep and the energy transfer to the lattice subsystem is predefined by the coupling factor and the temperatures of both subsystems, Langevin thermostat is well-suited for this situation \cite{yuan2021ablation}.
% The Langevin force was used to describe electron-phonon coupling in a stochastic way, modified based on Duffy and Rutherford’s works~\cite{duffy2006including, Rutherford_2007}, given by:
The electron-phonon coupling is therefore represented through a stochastic Langevin force ~\cite{duffy2006including, Rutherford_2007}:
\begin{equation}
    \label{F_langevin}
    \mathbf{F}_{\text{lang}} = -\frac{m_i G_{\text{e-ph}}(T_\text{e})}{3 n_i k_\text{b}} \mathbf{v}_T + \sqrt{2 k_\text{b} T_\text{e} \frac{m_i G_{\text{e-ph}}(T_\text{e})}{3 n_i k_\text{b}} \delta t} \, \bm{\beta}.
    % \tag{2.5}
\end{equation}
% \begin{equation}
%     \label{F_langevin}
%     \mathbf{F}^{\text{lang}}_j = -\frac{m_j G_{e\text{-}ph}(T_e)}{3 n_i k_b} \mathbf{v}_{T,j} + \sqrt{2 k_b T_e \frac{m_j G_{e\text{-}ph}(T_e)}{3 n_i k_b} \delta t} \, \bm{\beta}_j.
% \end{equation}
Here $m_i$ is the ionic mass, $n_i$ is the ionic density of the MD cell, $k_\text{b}$ is the Boltzmann constant, $\delta t$ is the timestep used in MD, $\bm{\beta}$ is a vector in three dimensions whose components are independent random variables, each having zero mean and unit standard deviation, and $\mathbf{v}_T = \mathbf{v} - \mathbf{v}_\text{c}$ with $\mathbf{v}$ and $\mathbf{v}_\text{c}$ being the velocity of the atom and the center of mass velocity of the MD cell, respectively.

The laser source term $\mathcal{P}$ in Eq.\eqref{eq:conduction_El_El_NEW} is calculated using FDTD simulations that solve nonlinear Maxwell equations and calculate the absorbed power density in the medium.

\subsection{Finite-Difference Time-Domain}

% FDTD directly integrates Maxwell’s curl equations in time and space by using the so-called Yee grid, in which the components of the electric and magnetic fields are staggered in both space and time \cite{yee1966,taflove2005}. The electric field is updated at integer multiples of discrete time steps, while the magnetic field is updated at half-integer multiples. These leapfrog updates are applied iteratively at each time step, advancing the fields through time.

The FDTD method directly integrates Maxwell’s equations in space and time \cite{yee1966, taflove2005}. We use the implementation tailored for nonlinear laser-matter interaction in structured materials \cite{fedorov2024light}.

The Maxwell’s curl equations are given as \cite{taflove2005}:
\begin{equation}
\begin{gathered}
\nabla \times \mathbf{E} = - \frac{\partial \mathbf{B}}{\partial t},\\
% - \mathbf{J}_B,
% \end{equation}
% \begin{equation}
\nabla \times \mathbf{H} = \frac{\partial \mathbf{D}}{\partial t}.
\end{gathered}
\end{equation}
where $\mathbf{E}$ is the electric field, $\mathbf{H}$ the magnetic field, $\mathbf{D}$ the electric displacement field, $\mathbf{B}$ the magnetic induction. The constitutive relations are $\mathbf{D} = \varepsilon_0 \varepsilon \mathbf{E} + \mathbf{P}$ and $\mathbf{B} = \mu_0 \mathbf{H}$, where
$\varepsilon_0$ is the vacuum permittivity,
$\mu_0$ the vacuum permeability,
and $\varepsilon$ the relative permittivity of the medium which represents all the linear nondispersive terms. Relative permeability $\mu$ is equal to 1 for silica.

Polarization $\mathbf{P}$ includes all the dispersive terms and nonlinearities relevant at high laser intensities:
\begin{equation}
\begin{gathered}
\mathbf{P} = \mathbf{P}_\text{disp} + \mathbf{P}_\text{nl} + \mathbf{P}_\text{p} + \mathbf{P}_\text{a},\\
\mathbf{P}_\text{disp} = \varepsilon_0 \chi^{(1)}_\text{disp}(\omega) \mathbf{E}, \\
% \mathbf{P}_{lin,nd} = \varepsilon_0 \varepsilon^{(1)} E, \\
\mathbf{P}_\text{nl} = \varepsilon_0 \chi^{(2)} \mathbf{E}^2 + \varepsilon_0 \chi^{(3)} \mathbf{E}^3, \\
\end{gathered}
\end{equation}
where $\mathbf{P}_\text{disp}$ is a dispersive polarization from dispersive materials such as Lorentz media \cite{taflove2005}, $\mathbf{P}_\text{nl}$ is the nonlinear term, and $\mathbf{P}_\text{p}$ and $\mathbf{P}_\text{a}$ are auxiliary polarizations defining the Drude plasma and the multiphoton absorption, respectively.

The plasma current density $\mathbf{J}_\text{p}$ follows directly from the time derivative of the plasma polarization $\mathbf{P}_\text{p}$: $\mathbf{J}_\text{p} = \frac{\partial \mathbf{P}_\text{{p}}}{\partial t}$. It is the continuous-time Drude current driven by the field in a collisional electron gas, and can thus be given as:
\begin{equation}
\frac{\partial^2 \mathbf{P}_\text{p}}{\partial t^2}
\;+\;
\nu_\text{c}\,\frac{\partial \mathbf{P}_\text{p}}{\partial t}
\;=\;
\frac{e^2}{m_\text{e}}\,n_\text{e}\,\mathbf{E},\\
\label{eq:drudeP}
\end{equation}
where $e$ is the electron charge, $m_\text{e}$ the electron mass, $\nu_\text{c}$ the electron collision frequency, and $n_\text{e}$ the free-electron density.

The multiphoton ionization current can be given via $\mathbf{P}_\text{a}$ as:
\begin{equation}
\frac{\partial \mathbf{P}_\text{a}}{\partial t} = K \hbar \omega_0\;\frac{\partial n_\text{e}}{\partial t}\;\frac{\mathbf{E}}{|\mathbf{E}|^2}\\
\end{equation}
where $K$ is the number of photons absorbed, and $\omega_0$ is the laser pulse central frequency.

The photoexcited electron density evolution is obtained from a nonlinear rate equation including Keldysh photoionization and avalanche ionization (field-driven impact ionization): 
\begin{equation}
\frac{\partial n_\text{e}}{\partial t} \;=\; W_{\text{pi}}(I)\,(n_{0}-n_\text{e}) \;+\; R_{\text{ai}}\, \mathbf{|E|}^2\,n_\text{e} ,\\
\label{eq:rho-ode}
\end{equation}
where $n_0$ is the initial density of valence electrons, and $R_{\text{ai}}$ is written in terms of a collisional absorption cross-section and the
bandgap $U_i$,
\begin{equation}
R_{\text{ai}} \;=\; \frac{\sigma_\text{B}}{U_{i}},
\qquad
\sigma_\text{B} \;=\; \frac{q_\text{e}^2}{m_\text{e}}\,\frac{\nu_\text{c}}{\nu_\text{c}^2+\omega_\text{0}^2},
\end{equation}
with $\sigma_\text{B}$ being the inverse Bremsstrahlung absorption cross-section.

\begin{figure}[!t]
\centering
\includegraphics[scale=0.275]{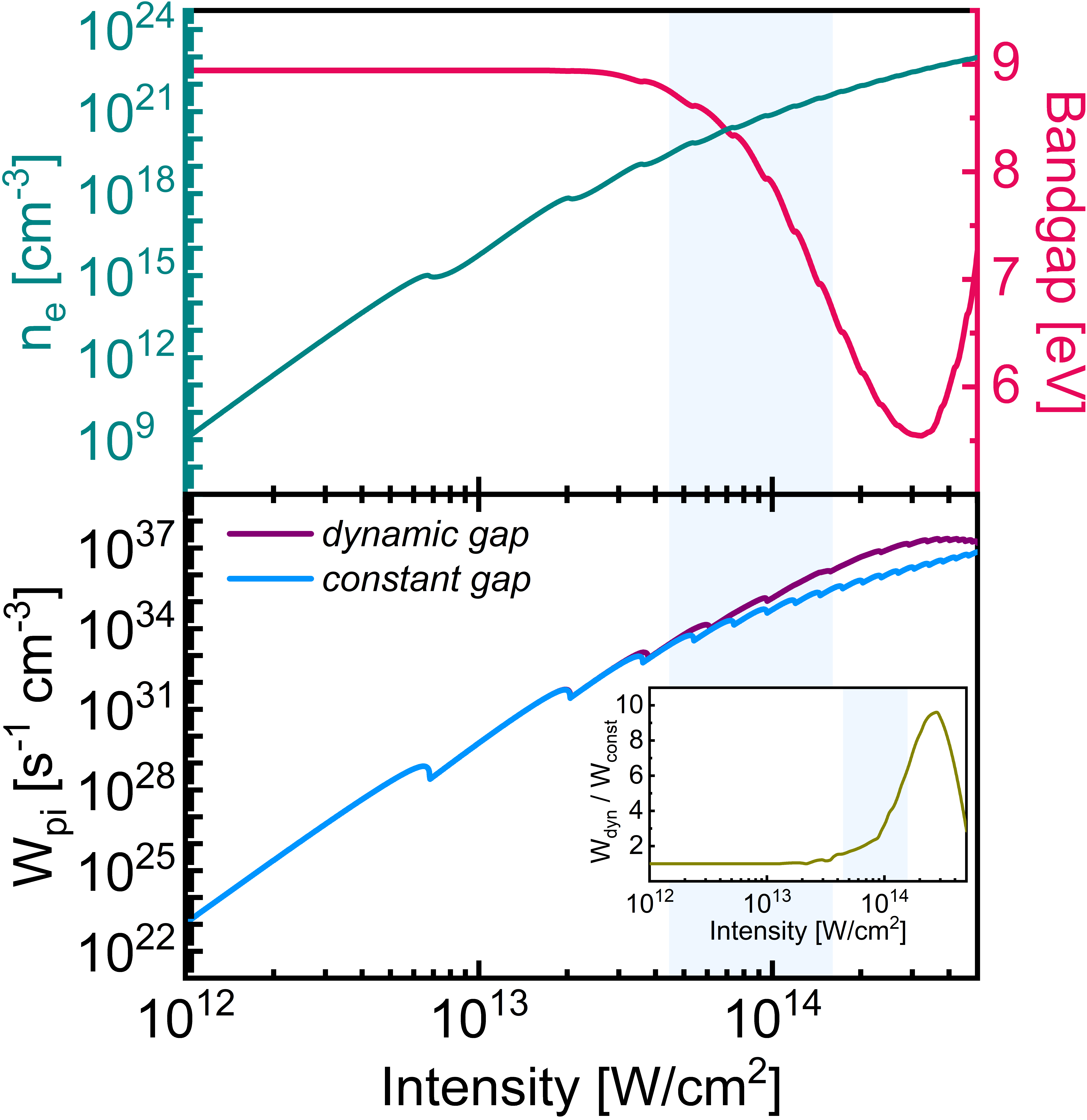}
\caption{(top panel) Free-electron density calculated with Keldysh theory generated by a laser pulse with $\lambda=1030$ nm and $\tau=25$ fs as a function of its peak intensity and the bandgap modified due to such density taken from Ref.~\cite{tsaturyan2024unraveling}. (bottom panel) Photoionization rate calculated with Keldysh theory with the dynamic bandgap taken from the top panel and constant bandgap (8.9 eV). The inset in bottom panel shows the smoothened ratio $W_\text{dyn}/W_\text{const}$ between ionization rates for dynamic and constant bandgap. The light gray area indicates the intensity range observed in FDTD simulations.}
\label{ionization_rate}
\end{figure}

Figure~\ref{ionization_rate} shows the photoionization rate $W_\text{pi}$ obtained from Keldysh formalism taking into account bandgap narrowing as a function of excited-electron density. The free-electron density generated by a femtosecond laser pulse is obtained by integrating the Keldysh photoionization rate over the temporal intensity profile of the pulse and then the bandgap is updated according to the Density-Functional Theory (DFT) results~\cite{tsaturyan2024unraveling} for the given laser pulse intensity. At high excited-electron density, the bandgap of silica is shown to decrease from 8.9 to 5.5 eV. As the bandgap narrows, the photoionization rate increases by up to an order of magnitude since fewer photons are required to cross the bandgap.

By denoting $\tilde{\mathbf D} = \mathbf D - \mathbf P_{\text{disp}} - \mathbf P_\text{p} - \mathbf P_\text{a}$, we can give the instantaneous isotropic or diagonal anisotropic nonlinear effects of the form:
\begin{equation}
\begin{gathered}
\tilde{D}_i \;=\; \varepsilon_0\,\varepsilon \, E_i \;+\; \varepsilon_0\,\chi^{(2)}_{i}\,E_i^{2} \;+\; \varepsilon_0\,\chi^{(3)}_{i}\,| \mathbf{E}|^2 E_i, \\
\end{gathered}
\end{equation}
Instead of implementing this equation directly, which would require solving a cubic equation at each time step, the code uses the following Padé approximation which matches the cubic solution to high-order accuracy \cite{oskooi2010}:
\begin{equation}
E_i \;=\;
\left[
\frac{1 + \dfrac{\varepsilon_0\,\chi^{(2)}}{[\varepsilon_0\,\varepsilon]^2}\,\tilde D_i
      + 2\dfrac{\varepsilon_0\,\chi^{(3)}}{[\varepsilon_0\,\varepsilon]^3}\,\Vert\tilde{\mathbf D}\Vert^2}
     {1 + 2\dfrac{\varepsilon_0\,\chi^{(2)}}{[\varepsilon_0\,\varepsilon]^2}\,\tilde D_i
      + 3\dfrac{\varepsilon_0\,\chi^{(3)}}{[\varepsilon_0\,\varepsilon]^3}\,\Vert\tilde{\mathbf D}\Vert^2}
\right]
\,[\varepsilon_0\,\varepsilon]^{-1}\,\tilde D_i .
\label{eq:meep-pade}
\end{equation}

In Maxwell’s equations, the power density $\mathcal{P}$ drawn from the field to the medium can be given through the plasma current by $\mathcal{P} \,=\, \mathbf{J}_\text{p}\,\!\cdot\,\!\mathbf{E}$.
This power density $\mathcal{P}$ is inserted as the source term in Eq.\eqref{eq:conduction_El_El_NEW}.

\subsection{Setup}

To model the post-compression state of silica following laser-induced shock wave propagation, we initialize our amorphous system at a density of $\sim$4.3 g/cm$^3$, corresponding to stishovite density at elevated pressure. This state represents shock-compressed amorphous silica immediately prior to crystallization, bypassing the initial compression phase which occurs on time and length scales beyond current MD capabilities \cite{noorArxiv, gleason2015ultrafast, grujicic2015densification}.
The dense amorphous silica is obtained from stishovite (lattice parameters $a = 4.17~\text{\AA}, c = 2.66~\text{\AA}$) by heat-anneal-quench-relax protocol using a supercell of 1350 atoms (Fig. S1 \cite{supMat}). The stishovite is heated to $8000~\mathrm{K}$ to ensure complete amorphization while maintaining high density, annealed for $0.5~\mathrm{ns}$, then quenched back to $300~\mathrm{K}$, followed by a final relaxation phase for another 0.5 ns.
The resulting structure is then replicated and relaxed to generate a larger supercell, with a volume of a few hundred cubic nanometers, containing a nanopore at its center. This dense amorphous nanocuboid with a pore in the center is our primary TTM-MD system subjected to laser irradiation. We consider nanopores with two diameters of 2 and 4 nm.

The resolution used for FDTD simulations is 0.125 nm. The absorbed laser power distributions obtained in FDTD are then interpolated to the MD resolution, which in our simulations is 0.95 nm. This is the lowest resolution for TTM-MD simulations to avoid incorrect energy gradients, over/under-heated cells, wrong dynamics and simulation crash. The refractive index used is $n = 1.45$. The neutral electron density is $n_\text{e} = 3.5 \times 10^{23}$ cm$^{-3}$, and the collision frequency is $\nu_\text{c} = 4.29 \times 10^{13}$ s$^{-1}$ \cite{sudrie2002femtosecond}. The reduced effective mass used for Keldysh theory is 0.64$m_\text{e}$ \cite{couairon2005filamentation}. Since the structure length (8 nm) is very small compared to wavelength, the dispersion is negligible ($\chi_{\mathrm{disp}}^{(1)}(\omega)= 0$). $\chi^{(2)}=0$ due to centrosymmetry in silica, and for the third-order susceptibility we take $\chi^{(3)} = 2.5\times 10^{-22}\ \mathrm{m^2/V^2}$ \cite{boyd2008nonlinearOptics}. The structure is irradiated by a Gaussian laser pulse with peak intensity of $I=10^{14}$ W/cm$^2$, wavelength of $\lambda=1030$ nm, and pulse duration of $\tau=25$ fs at full-width half-maximum (FWHM).

The electronic specific heat $C_\text{e} (T_\text{e})$ and electron-phonon coupling factor $G_{\text{e-ph}} (T_\text{e})$, used in Eq.\eqref{eq:conduction_El_El_NEW}, are taken from a recent study by Tsaturyan \textit{et al.} which investigates ultrafast laser excitation of $\alpha-$quartz by applying high electron temperature according to Fermi-Dirac distribution of electrons using DFT, GW approximation, and density functional perturbation theory (DFPT) \cite{tsaturyan2024unraveling}.
The electronic thermal conductivity $K_\text{e} (T_\text{e})$ is calculated as in Ref. \cite{burakov2005theoretical}.

% Although the mass densities and long-range order differ in $\alpha-$quartz and high-density amorphous states, to our knowledge there is no study that reports a complete, self-consistent set of electronic parameters over electronic temperature at different, especially high densities. Moreover, the existing studies do not take into account the band-gap evolution at high intensities which is crucial for FDTD simulations in inhomogeneous media. Consequently, adopting the dataset discussed above for our model is a reasonable approximation.

The modified BKS (Beest-Kramer-Santen) interatomic potential is used in MD \cite{PhysRevE.63.011202}. The MD simulations are performed by LAMMPS \cite{thompson2022lammps}. Visualization is performed by OVITO \cite{ovito}.

\section{Results}

\begin{figure*}[!t]
\centering
\includegraphics[scale=0.191]{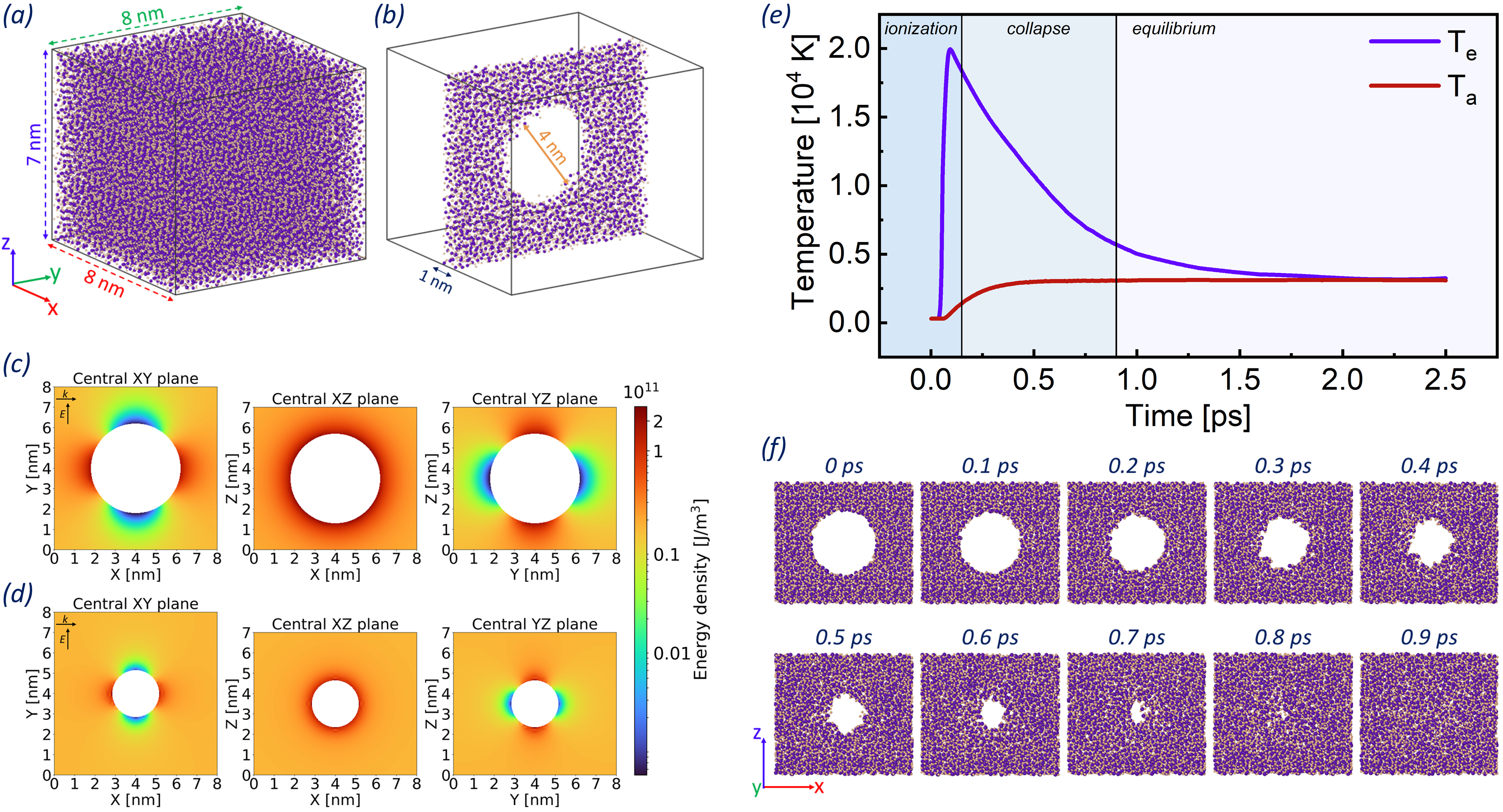}
\caption{(a) Amorphous silica nanocuboid used in the simulation. 
(b) Central 1-nm-thick slice of the 4-nm-pore system along the $x$-direction, corresponding to a 7\% porosity for the simulation cell volume of $\sim$450 nm$^3$.
(c, d) Central slices of the absorbed energy density distributions obtained from FDTD simulations for 4-nm- and 2-nm-pore systems, respectively.
The laser wavelength, pulse duration, and peak intensity are 1030 nm, 25 fs, and $10^{14}$ W/cm$^2$, respectively.
(e) Temporal evolution of the average electronic and atomic temperatures during the first few picoseconds following laser excitation. 
(f) Snapshots of a central 1-nm-thick slice of the simulation box during the first picosecond, showing the onset of laser-driven nanopore collapse.}
\label{Mixed}
\end{figure*}

The dense amorphous nanocuboid with a nanopore in the center, used in MD simulations, and its 1-nm-thick central slice are shown in Figs.~\ref{Mixed}(a-b), respectively. The absorbed laser energy distributions obtained by FDTD for structures with 4- and 2-nm pores are shown in Figs.~\ref{Mixed}(c-d), respectively.
The laser pulse is linearly polarized along Y- and propagates along X-direction, leading to a pronounced anisotropic absorption pattern. Because the electric field oscillates along Y, the tangential field components remain continuous across the interface, while the normal component of the displacement field must satisfy boundary conditions that enhance the electric field inside the high-index region near the pore boundary. This mismatch in permittivity produces localized field amplification, particularly at the equatorial regions of the pore aligned with the polarization axis. The amplified local-fields and, thus, the local intensities contribute through the nonlinear photoionization response to the generation of highly concentrated free-electron densities and absorption hotspots. These electromagnetic hotspots are the primary source of spatially localized energy deposition that seeds subsequent thermomechanical evolution. The spatial distributions of free-electron density, laser intensity, and bandgap are provided in Figs. S2-S3 \cite{supMat}.
% The energy concentration in these planes confirms the expected polarization-induced enhancement, and supports the use of this energy profile as the source term in the coupled TTM-MD model.

Figure~\ref{Mixed}(e) presents the temporal evolution of the 4-nm-pore system average electronic temperature $T_\text{e}$ and atomic (lattice) temperature $T_\text{a}$ after femtosecond laser irradiation.
The electronic temperature $T_\text{e}$ peaks at 75 fs (the peak of the laser pulse) to nearly 20000 K.
Then, via fast electron-phonon coupling, the hot electrons transfer their energy to the lattice within several picoseconds, raising it to 3100 K.
Such high temperature leads to the pore collapse.

% The temperature evolution also explains the dynamics of the pore collapse observed in the snapshots in Fig. \ref{Mixed}(f). The snapshots show the dynamics of the pore of a 1-nm thick central slice during the first picosecond following the ultrafast energy deposition.
Due to enhanced laser-energy absorption around the pore, the surrounding silica matrix experiences a rapid increase in temperature and pressure. This creates a steep pressure gradient between the heated shell around the pore and the pore interior. As a result, atoms from the high-pressure shell begin to move inward, leading to the collapse of the pore within 1 ps (Fig. \ref{Mixed}(f)). At this stage, the structure is highly inhomogeneous in terms of temperature, pressure and density.
This behavior is characteristic of thermomechanical response under extreme non-equilibrium conditions, where the fast conversion of absorbed optical energy into kinetic motion drives rapid material flow into the pore. The presence of strong heterogeneous temperature and pressure around the pore can also accelerate the collapse via localized momentum transfer. 

The lattice heating is weaker in the case of smaller 2-nm-pore (final temperature of 2670 K) and homogeneous medium (final temperature of 2580 K) because the local-field enhancement depends on the pore size. The laser energy absorption is higher around the larger 4-nm-pore that leads to a higher lattice temperature of 3100 K. The small temperature difference between the 2-nm and homogeneous systems indicates that sub-nanometer porosity contributes only marginally to energy localization under these irradiation conditions. Since the equilibrium temperatures of the 2-nm-pore and homogeneous systems remain well below the threshold required for a phase transition at the given density, no structural transformation is observed even after several nanoseconds. Below, only the 4-nm-pore system, where the transition occurred, is discussed in detail.

\begin{figure*}[!t]
    \centering
    \includegraphics[scale=0.3]{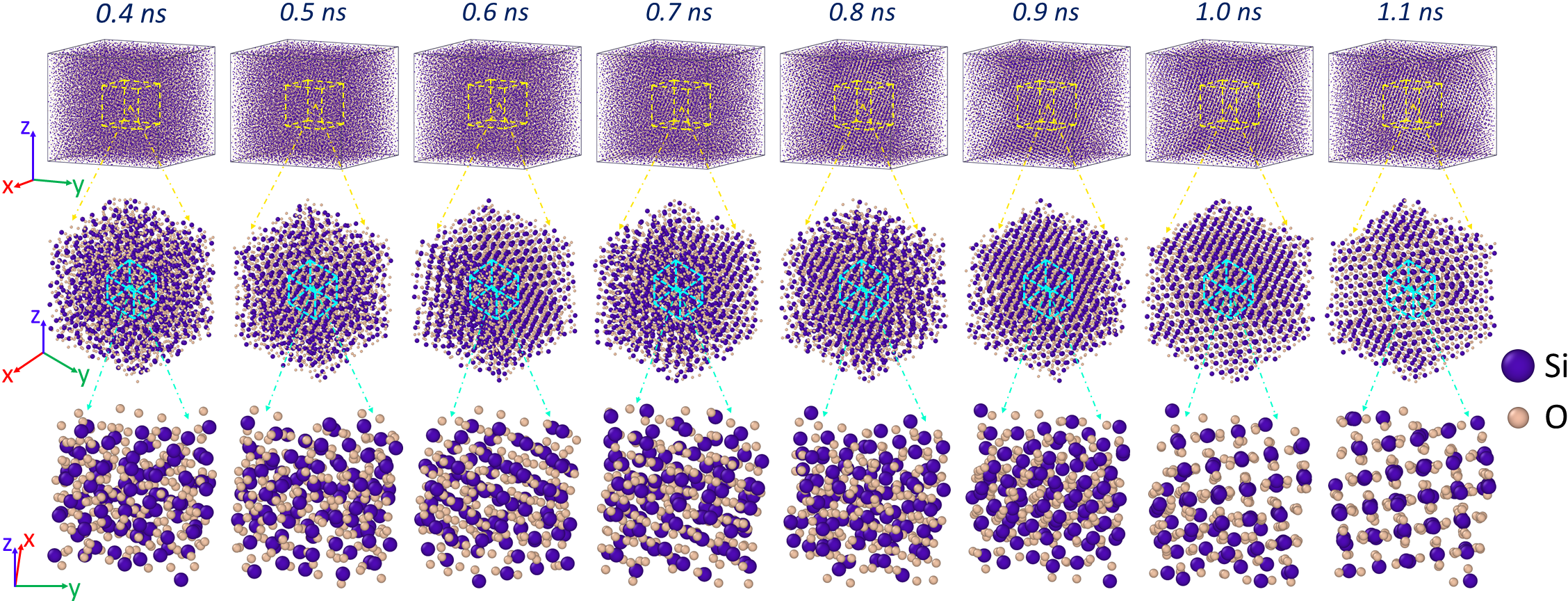}
    \caption{Snapshots of the simulation box at successive times during crystallization. The first row shows the full simulation box, the second row a central cuboid region of approximately 3 nm per side, and the third row a smaller subsection extracted from this region to highlight the emerging crystalline order.}
    \label{snapshots_crystallization}
\end{figure*}

Figure \ref{snapshots_crystallization} shows snapshots of the atomic configuration during the crystallization stage for the full simulation box and central zoomed areas.
The first crystalline nuclei appear around 0.6 ns as seen in zoomed areas. The crystal then grows further and the entire structure crystallizes within half a nanosecond.
The system is largely transformed into the crystalline stishovite phase, although structural relaxation continues for several tens of nanoseconds as the lattice progressively minimizes its energy. Minor differences between atomic coordinates in neighboring snapshots arise from atomic motion across periodic boundaries.

\begin{figure*}[!t]
\centering
\includegraphics[scale=0.671]{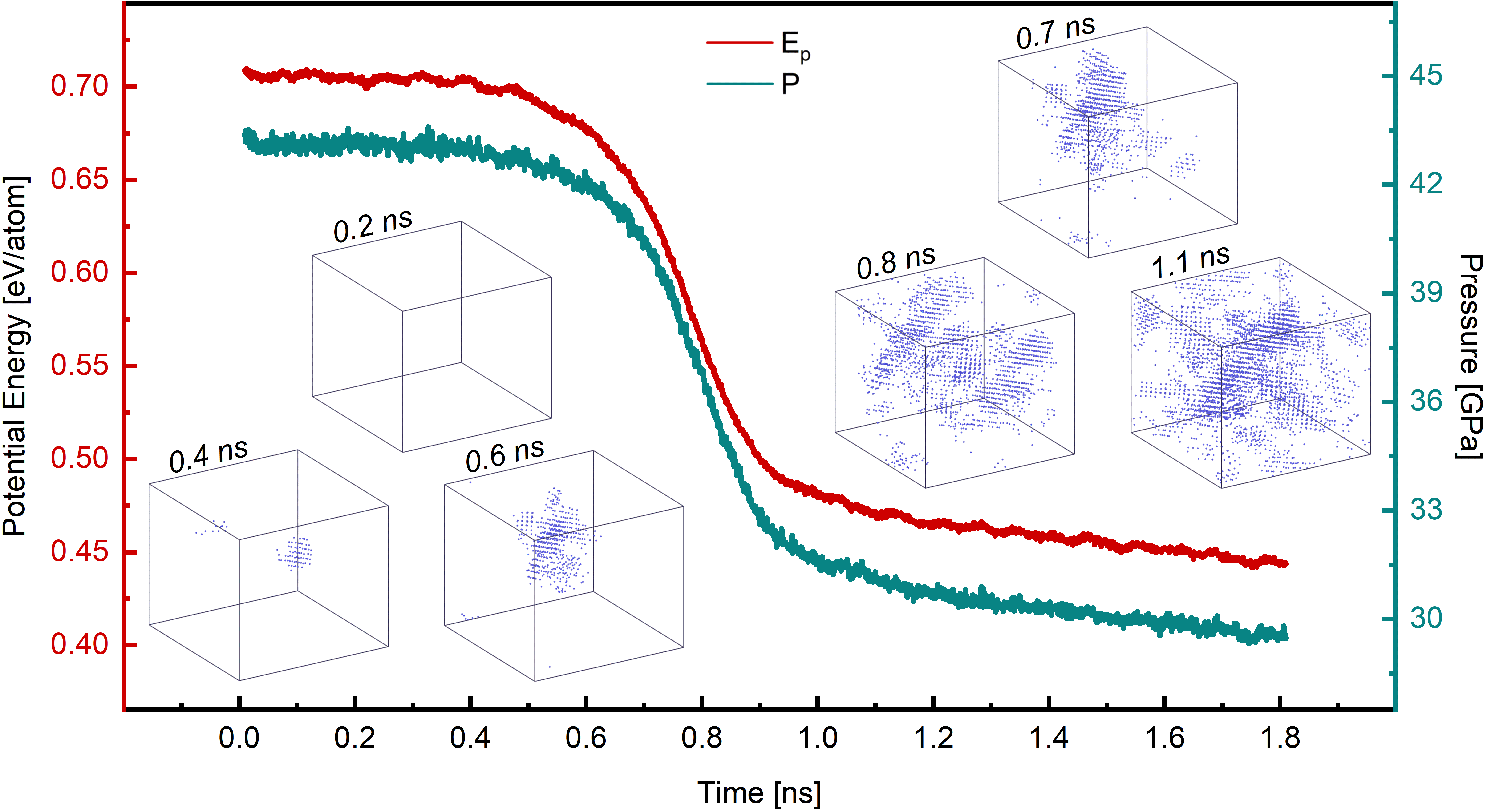}
\caption{Temporal evolution of the potential energy and pressure after thermal equilibrium. The decrease of both quantities indicates the onset of the phase transition. Insets show atomic snapshots illustrating the nucleation and growth of stishovite-like clusters (Si atoms only).}
\label{Ep_and_P}
\end{figure*}

The nucleation and crystallization process is further explored by examining the system potential energy and pressure evolution (Fig. \ref{Ep_and_P}). The potential energy and pressure start decreasing at around 0.35 ns, marking the onset of critical nucleus formation which leads to crystallization. Around 0.5 ns the decrease in both quantities becomes abrupt, a characteristic signature of a first-order phase transition. At this point the system crosses the nucleation threshold and enters a fast transformation toward the stishovite phase. The sharp drop of potential energy by $\sim$0.27 eV/atom is accompanied by a pressure drop of $\sim$14 GPa, consistent with the collapse of the disordered structure into an energetically more favorable configuration. This transition occurs within a narrow time window of around 0.5 ns, signifying that crystallization proceeds rapidly once a critical nucleus forms. After around 1.1 ns the decrease of potential energy and pressure slows down, indicating that the phase transition is completed, which closely matches the timeframes of the crystallization snapshots shown in Fig. \ref{snapshots_crystallization}. However, the decrease in potential energy and pressure still continues during tens of nanoseconds until the system finds the minimum, energetically most favorable state.

\begin{figure}[!t]
\centering
\includegraphics[scale=0.34]{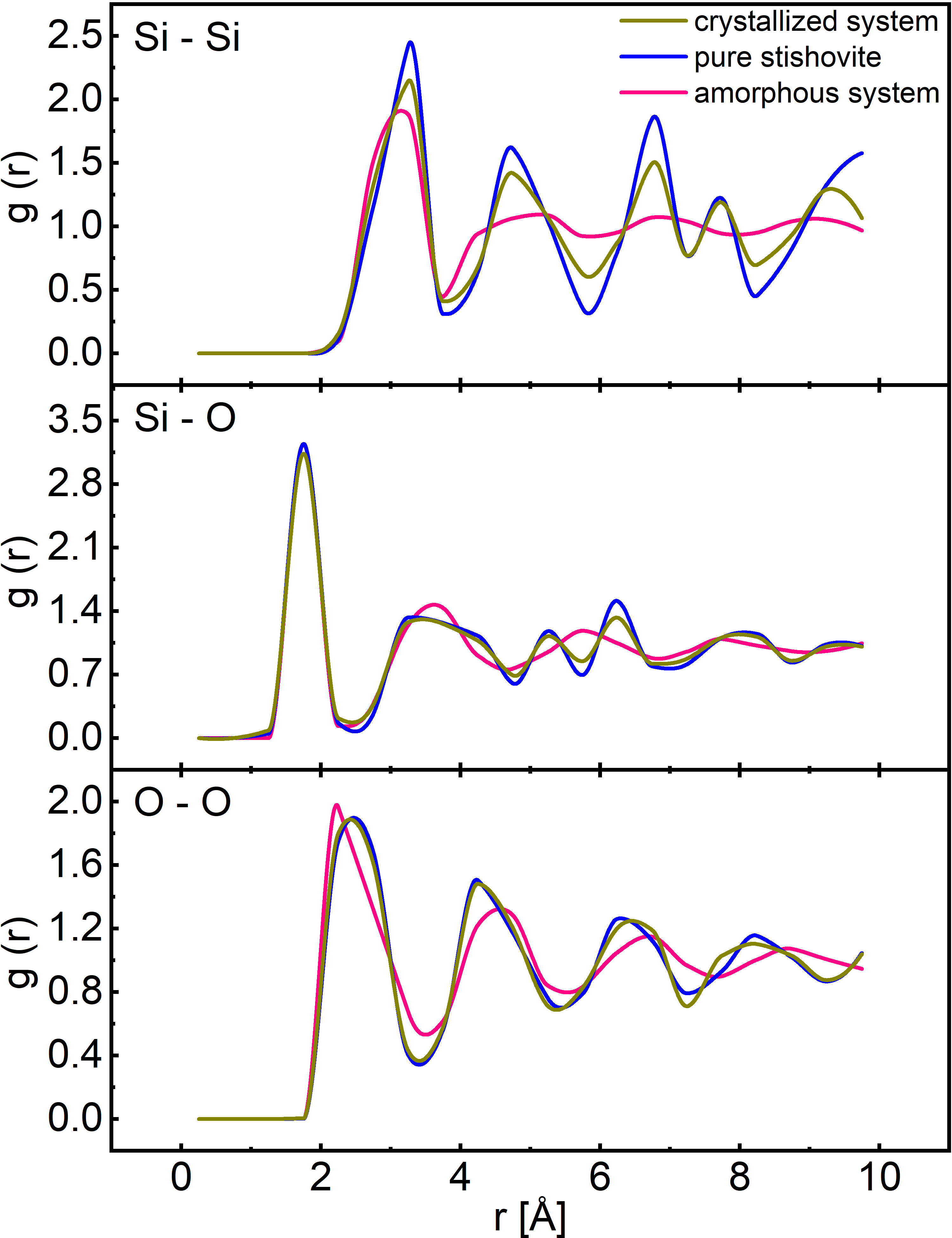}
\caption{Partial radial distribution functions for Si–Si, Si–O, and O–O atomic pairs in the crystallized system, compared with those of pure stishovite and amorphous silica under identical thermodynamic conditions, confirming the structural signature of the stishovite phase.}
\label{RDF_all}
\end{figure}

The partial radial distribution functions (PRDFs) for Si-Si, Si-O, and O-O atomic pairs of final crystal structure are shown in Fig.~\ref{RDF_all}. The PRDFs of pure stishovite and of the initial amorphous structure are shown for comparison. The pure stishovite crystal and initial amorphous structure are heated to the same temperature to take into account the thermal noise effects when comparing the PRDFs. The peaks' positions for all atomic pairs are almost the same for the crystallized system and pure stishovite. This close agreement provides strong evidence that the crystallized system has the stishovite phase. In contrast to the crystalline phase, the PRDFs of the amorphous structure exhibit significantly broader and less pronounced peaks. This reflects the lack of long-range periodicity and the presence of local structural disorder. While the first-neighbor peak is still visible, indicating short-range order, the higher-order coordination shells are smeared out or entirely absent, consistent with the random network topology of amorphous silica.

While the PRDFs of the crystallized and pure stishovite systems match perfectly for Si-O and O-O pairs, the peaks' amplitudes are slightly different for the Si-Si pair. The root mean square deviation (RMSD) between the PRDFs is 8\% for Si-O and 5\% for O-O pairs indicating a strong preservation of the local bonding environment. The first-neighbor peaks in all PRDFs are in excellent agreement, suggesting minimal disruption of short-range order.
In contrast, the Si-Si PRDF exhibits a notably higher RMSD of 23\%. While the first two Si-Si peaks, representing the immediate and second Si neighbors, align well in both position and intensity, discrepancies emerge in the subsequent coordination shells. These deviations are not due to changes in bond distances alone but are instead attributed to distortions in medium- and long-range order, discussed in detail in the following section.

\section{Discussion}

There are several reasons for the distortions in Si-Si pairs. First, the distorted structure has the stishovite topology but with a reorientation of its crystallographic axes. Specifically, the c-axes of the crystallites in the distorted structure no longer align with the original [001] direction of the simulation box but are instead oriented along a nontrivial direction, forming oblique angles with the simulation cell axes. This crystallographic misalignment, which likely arises from random nucleation events, alters the spatial distribution of interatomic distances beyond the first coordination shell, leading to broadened or asymmetric RDF peaks, despite their positions remaining equal to those of the reference.

Second, the crystallized system contains approximately 6\% fewer atoms than the reference crystal because it originates from a nanoporous system, whereas the reference crystal is obtained through standard unit-cell replication. This difference affects the normalization of the RDF and introduces amplitude discrepancies, particularly for Si-Si pairs, which are more sensitive to system size and connectivity.

Third, the presence of a nanopore inevitably perturbs the local stoichiometry. This imbalance, combined with independent nucleation effects, can lead to the formation of structural defects such as interstitials, Frenkel pairs, stacking faults, and grain boundaries. These imperfections are particularly disruptive to the long-range order probed by the Si-Si PRDF, where coordination shells depend on precise lattice periodicity. Even in perfectly stoichiometric systems, defects may still arise from the convergence of multiple growing crystallites, whose random orientations and mismatch at interfaces produce disorder beyond the first few atomic shells.

Finally, stishovite is known to undergo a transition to the orthorhombic CaCl$_2$-type post-stishovite phase at higher pressure and temperature, which preserves the sixfold coordination of silicon but involves cooperative tilting and distortion of the SiO$_6$ octahedra, lowering the crystal symmetry from tetragonal to orthorhombic \cite{pan}. The coexistence of regions with different distortion states or local transition-induced strains can further disrupt medium- and long-range structural order, broadening the features observed in the Si-Si PRDF.

The local structural motifs of the crystallized phase were further characterized through bond-angle and bond-length distributions. The obtained O-Si-O and Si-O-Si bond-angle distributions (Fig. S4 \cite{supMat}), as well as the Si-O bond-length distribution (Fig. S5 \cite{supMat}), are consistent with the octahedral coordination and connectivity expected for stishovite and are in good agreement with previous experimental and computational studies \cite{nhan_2019, misawa_2017, PhysRevB.82.184102}.

We performed additional simulations where the laser pulse intensity was adjusted in order to bring the systems with 4-nm pore, 2-nm pore and without pore to the same final temperature (3100 K) and average density (4.35 g/cm$^3$) after equilibrium. Interestingly, even under such equivalent thermal load, crystallization is almost twice as slow in the homogeneous case than in the porous structure. The 4- and 2-nm-pore structures crystallize after about 0.8 ns, whereas the pore-free structure crystallizes 1.4 ns after irradiation.

This delay can be attributed to the absence of pre-existing inhomogeneities, which otherwise act as natural sites for energy localization and nucleation.
Nanopores introduce localized regions of high pressure and temperature gradients, due to the laser energy concentration around the pore and its subsequent rapid collapse. These gradients create favorable conditions for atomic rearrangements and drive the system out of equilibrium promoting faster formation of critical nuclei.
In a structurally uniform medium, the system must rely on spontaneous thermal fluctuations to initiate crystallization. This process is statistically less probable and slower, resulting in a later phase transformation. These findings support the broader view that localized structural or compositional inhomogeneities, such as pores, density fluctuations, or bond distortions, can significantly reduce the energy barrier for nucleation, thereby accelerating the overall crystallization process.
% The simulation results presented here strongly support the conclusion that the presence of nanopores significantly accelerates the crystallization process in laser-heated silica. In our setup, where the sample reaches an equilibrated temperature of approximately 3100~K, a marked difference in the onset of crystallization is observed depending on the presence and size of the pore. For instance, when a 2~nm radius void is introduced at the center of the simulation cell, the system exhibits a slow decline in pressure and potential energy around 0.35~ns, followed by a rapid drop corresponding to a crystallization event. Interestingly, this timeline remains nearly unchanged even when the pore radius is reduced twice. However, in a fully homogeneous structure, the critical nucleus formation is delayed significantly, occurring around 0.9~ns.
The nucleation, however, appears insensitive to pore size within the 4-nm range, suggesting that even minimal nanoscale inhomogeneities are sufficient to trigger this effect.
Whether this behavior extends to pores with radii significantly larger than those considered here, where altered field enhancement, pressure confinement, or thermal diffusion may modify nucleation kinetics, remains unexplored.
% In classical nucleation theory \cite{becker1935kinetische}, the nucleation rate depends exponentially on the free-energy barrier for forming a critical nucleus, which reflects the competition between interfacial energy and the thermodynamic driving force. Temperature affects the rate through thermal activation and its influence on both the driving force and kinetic factors. The presence of inhomogeneities, such as pores, interfaces, or compositional gradients, can significantly reduce the nucleation barrier accelerating the process.

Moreover, the collapse of the pore itself is a strongly nonequilibrium process that also leads to a localized spike in pressure and temperature.
These phenomena, enhanced by the spatial inhomogeneities, indicate that localized structural fluctuations, like coordination number distributions, broken rings or trapped molecular species, could substantially modify the thermodynamic and kinetic landscape of crystallization.

As confirmed by the evolution of the potential energy, the phase transition appears to be a two-stage process: first, the formation and stabilization of critical nuclei, followed by a collective phase transformation. These observations provide insight into how nanostructuring can be used to tune crystallization kinetics in experimental settings.

This behavior can be rationalized in light of our experimental observations presented in Sec. \ref{sec:experiments}, which reveal that most crystallized regions in laser-irradiated silica glass are localized near interfaces rather than in the bulk. These interfacial zones exhibit enhanced thermal and pressure gradients, localized density fluctuations, and geometric complexity. Such features can locally reduce the nucleation energy barrier by altering the free energy landscape, thereby facilitating the crystallization process.

% In addition to spatial inhomogeneities, the dynamics of cooling plays a crucial role. In laser-irradiated systems, the heating phase is extremely rapid, driven by femtosecond to picosecond energy deposition, while the subsequent cooling is significantly slower. In MD simulations, it is computationally infeasible to model realistic cooling rates, which would occur over nanoseconds to microseconds in experiments. In our model, the simulation box is small and positioned at the center of the laser pulse, where the intensity is assumed to be maximal. Consequently, the rate of energy loss is minimized, and we simulate a quasi-isolated system in which most absorbed energy remains within the material. This setup allows us to observe crystallization under idealized, high-energy retention conditions, even though such behavior is more difficult to realize experimentally in homogeneous media.

Cooling dynamics also plays an important role in the crystallization process. After laser excitation, the material transitions into a high-temperature, high-pressure melt, which then cools down and reaches thermodynamic conditions favorable for nucleation for a limited time. In homogeneous regions, this time may not be sufficient to overcome the high nucleation barrier, so the system fails to form a critical nucleus and instead cools into a non-crystalline, glassy state. In contrast, interfacial or porous regions, owing to their lower effective nucleation barriers, are more likely to initiate crystallization within this limited time.

Therefore, both our simulations and experiments show that local structural inhomogeneities enhance energy localization and increase the nucleation probability, thereby triggering phase transitions.

\section{\label{sec:experiments}Experiments}

\begin{figure*}[!t]
\centering
\includegraphics[scale=0.9]{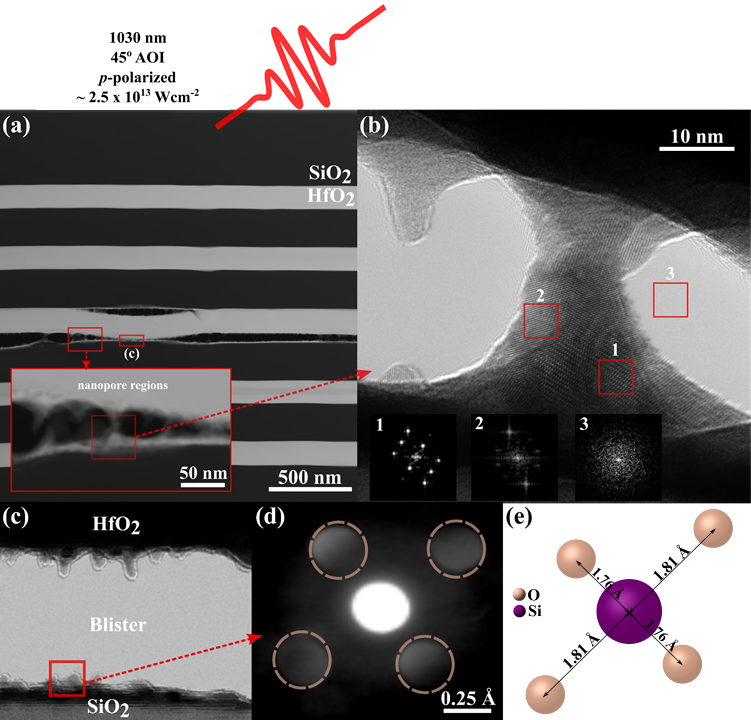}
\caption{Laser-induced nano structural modification and crystallization at the SiO$_2$/HfO$_2$ interface. (a) Cross-sectional TEM image of the SiO$_2$/HfO$_2$ MLD after single-shot femtosecond laser irradiation, showing localized interfacial deformation and blister formation. The inset shows a nanoporous region formed within the SiO$_2$ and HfO$_2$ layers adjacent to the interface, indicative of rapid volumetric expansion and material rarefaction.  (b) High-resolution TEM image of the boxed nanoporous region in (a), revealing nanometer-scale crystalline domains embedded within an amorphous matrix near the SiO$_2$/HfO$_2$ interface. The insets show the Fourier transforms of regions 1, 2 and 3. (c) 4D-STEM real-space scan across the modified interface, with the boxed region used for nano diffraction. (d) Nano-diffraction pattern extracted from the laser-modified region using 4D-STEM, showing four well-defined diffraction maxima arranged in a near-square geometry; overlaid guides indicate the measured reciprocal-lattice spacings. (e) Simulated stishovite structure and diffraction viewed along the c-axis; experimental spacings agree within 3\%, confirming formation of stishovite at the SiO$_2$/HfO$_2$ interface.}
\label{experiment1}
\end{figure*}

Amorphous SiO$_2$/HfO$_2$ multilayer dielectric (MLD) mirror samples are irradiated using a femtosecond laser pulse at 1030 nm wavelength and duration of 70 fs in a single-shot regime allowing to avoid cumulative effects. The complete laser system, irradiation geometry, damage protocol, surface-morphology characterization, and sample preparation procedures for transmission electron microscopy (TEM) are described in detail in our prior works \cite{noor2025pulseDuration, noorArxiv}. Site-specific cross-sectional lamellae from laser-modified (blistered) regions are prepared by focused ion beam milling with protective Pt capping and thinned to $\sim$100 nm for TEM analysis. Structural characterization is performed using bright-field and high-resolution TEM imaging, selected-area electron diffraction (SAED), and four-dimensional scanning transmission electron microscopy (4D-STEM). SAED patterns acquired from sub-micrometer regions are indexed by comparing measured $d$-spacings with simulated diffraction patterns, while 4D-STEM nano-diffraction datasets are obtained by raster-scanning a focused probe and recording a full diffraction pattern at each position, enabling spatially resolved phase identification within the laser-modified volume. 

Figure \ref{experiment1} illustrates the nanoscale structural evolution induced at the SiO$_2$/HfO$_2$ interface of the MLD mirror following femtosecond laser irradiation. Fig. \ref{experiment1}(a) is the cross-sectional TEM image of the laser-modified multilayer stack showing pronounced interfacial damage morphology. In addition to layer distortion and blister formation as reported in Ref. \cite{noor2025pulseDuration}, a nanoporous region is observed in the SiO$_2$ and HfO$_2$ layers, as highlighted in the inset in Fig. \ref{experiment1}(a). This region consists of nanoscale voids and density fluctuations extending laterally along the interface, indicative of rapid volumetric expansion and material rarefaction following ultrafast energy deposition. Such nanoporous features are consistent with strong-field ionization, confined plasma formation, and subsequent pressure-driven relaxation in wide-bandgap oxides. Notably, although most of the laser energy is absorbed in the first HfO$_2$ layer, the FDTD simulations show that appreciable free-electron densities are still generated within the third HfO$_2$ layer, where subsurface plasma confinement and interfacial stress localization favor blister-driven damage initiation beneath the surface \cite{noor2025pulseDuration}. Fig. \ref{experiment1}(b) is the high-resolution TEM image of the boxed nanoporous region in (a) which reveals localized nanometer-scale crystalline domains embedded within an otherwise amorphous matrix near the SiO$_2$/HfO$_2$ interface. The Fourier transforms of the regions 1 and 2 show discrete diffraction spots characteristic of crystalline order, while region 3 exhibits a diffuse halo typical of amorphous SiO$_2$, confirming the highly localized nature of laser-induced crystallization. To spatially resolve this structural modification with crystallographic sensitivity, a 4D-STEM real-space scan is acquired across the damaged interface, as shown in Fig. \ref{experiment1}(c), with the red inset marking the region selected for nano diffraction analysis. The corresponding diffraction pattern extracted from this region, shown in Fig. \ref{experiment1}(d), exhibits four pronounced diffraction maxima arranged in a nearly square geometry, signaling the emergence of long-range order within the laser-modified SiO$_2$ layer. Comparison with simulated diffraction from high-pressure SiO$_2$ polymorphs reveals excellent agreement with stishovite viewed along the crystallographic c-axis (Fig. \ref{experiment1}(e)). The experimentally measured diffraction vectors match the simulated stishovite reciprocal spacings within $\sim$3\%, confirming that femtosecond laser excitation locally transforms amorphous SiO$_2$ into crystalline stishovite at the SiO$_2$/HfO$_2$ interface. These observations demonstrate that femtosecond laser excitation induces extreme, highly localized pressure transients at the SiO$_2$/HfO$_2$ interface, leading to nanoporous void formation, interfacial delamination, and non-equilibrium crystallization of amorphous SiO$_2$ into the high-pressure stishovite phase, processes that are inaccessible under equilibrium thermodynamics conditions.

\begin{figure}[!t]
\centering
\includegraphics[scale=0.796]{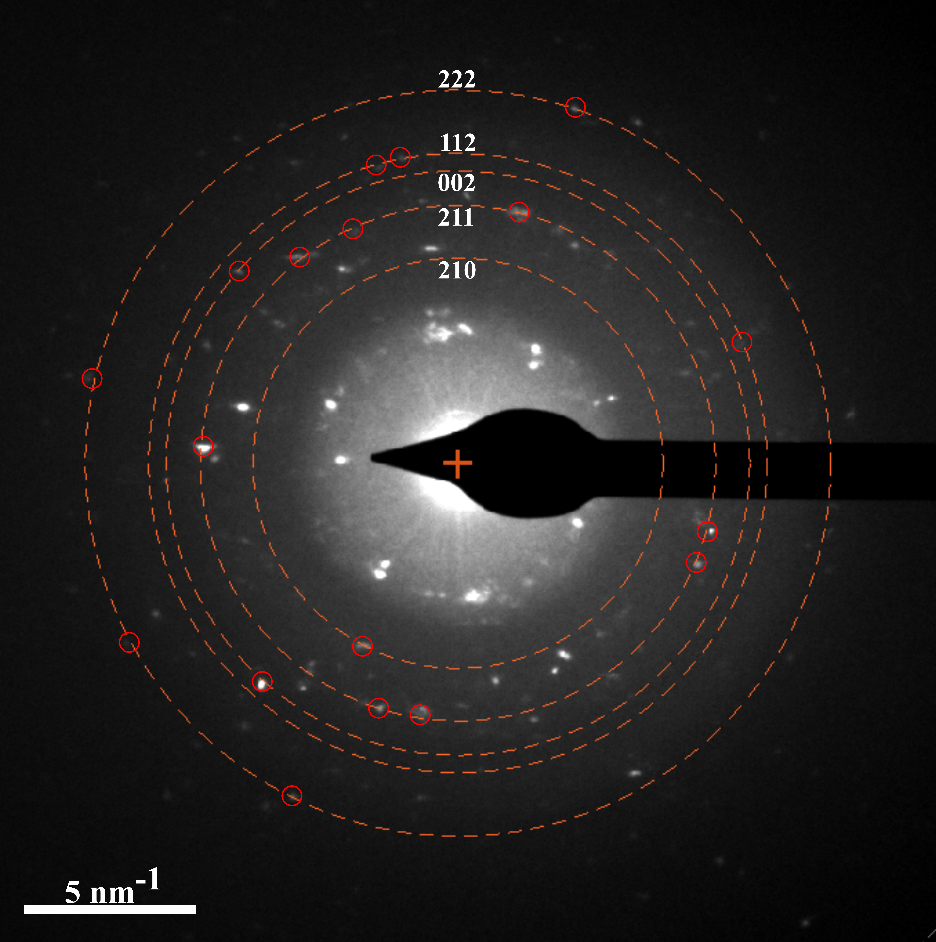}
\caption{SAED pattern of laser-induced high-pressure stishovite phases; red circles denote experimental spots, and dashed rings indicate indexed (hkl) planes.}
\label{experiment2}
\end{figure}

Figure \ref{experiment2} shows a representative SAED pattern acquired from the laser-modified SiO$_2$ region at the SiO$_2$/HfO$_2$ interface. The diffraction pattern consists of multiple discrete spots superimposed on a weak diffuse background, indicating partial crystallization within an amorphous matrix. The dashed concentric rings correspond to calculated reciprocal-lattice spacings of high-pressure SiO$_2$ polymorphs, while the indexed reflections (210), (211), (002), (112), and (222) match well with those expected for stishovite. The red-circled diffraction maxima align closely with the simulated ring positions, confirming the presence of nanocrystalline domains with no strong preferred orientation. The measured reciprocal spacings agree with simulated stishovite values within experimental uncertainty of 2\%, consistent with shock-driven crystallization under ultrafast laser-induced pressure transients.

While the present experimental study isolates single-pulse interactions, we believe that multipulse irradiation is expected to introduce additional and qualitatively different modification pathways. Repeated femtosecond excitation at the same site enhances crystallization by stabilizing laser-induced nuclei and lowering the effective nucleation barrier through defect accumulation, residual densification, and localized field enhancement at pre-existing nanopores and the SiO$_2$/HfO$_2$ interface. However, with increasing pulse number, cumulative void growth, pressure venting, and thermo-mechanical loading suppress the ultrafast shock confinement required for high-pressure phase formation, driving a transition from localized non-equilibrium crystallization (e.g., stishovite) toward defect-assisted amorphization, interfacial delamination, and eventual catastrophic material removal. A further study with different numbers of pulses, repetition rate and laser parameters is expected in a future work.

\section{Summary}

We show that nanoscale porosity strongly modifies the ultrafast response of silica to femtosecond laser irradiation and can decisively alter its crystallization kinetics. By combining nonlinear FDTD simulations of laser energy deposition with atomistic Two-Temperature Molecular Dynamics, we demonstrate that local-field enhancement at a nanopore interface concentrates electromagnetic energy and drives rapid pore collapse. The resulting transient pressure and temperature conditions enable the nucleation and growth of crystalline stishovite on sub-nanosecond timescales.
Depending on the laser intensity, crystallization in homogeneous systems occurs much later or not at all.

The simulations reveal that crystallization in porous silica is taking place while pressure is high enough, allowing the formation of stishovite before pressure relaxation suppresses the transition. In contrast, homogeneous systems exhibit a delayed onset of nucleation, despite reaching comparable average temperature and density. This behavior highlights the critical role of structural inhomogeneities in reducing nucleation barriers and promoting heterogeneous crystallization under strongly nonequilibrium conditions. The acceleration of crystallization is shown to be rather insensitive to pore size and can be triggered even by nanometer-scale pores.

These results provide a mechanistic explanation for experimental observations of laser-induced crystallization of high-pressure SiO$_{2}$ polymorphs at interfaces and in confined geometries by reproducing the same spatial selectivity and phase outcome. In particular, the formation of stishovite observed experimentally in laser-irradiated silica multilayers and near voided regions is consistent with the pore-mediated energy localization and pressure confinement observed in simulations. This work thus demonstrates that nanoscale porosity can be exploited as an effective control parameter for directing ultrafast solid-state transformations. The framework developed here is applicable to other transparent dielectrics and nanoporous materials and provides a predictive basis for engineering laser-driven synthesis of dense and metastable crystalline phases under far-from-equilibrium conditions. Future extensions of this framework could model laser-induced phase transitions in other wide-bandgap systems such as Al$_2$O$_3$, ZrO$_2$ and HfO$_2$, and establish a general computational tool for predicting and controlling laser-driven structural transformations in dielectric materials.

\begin{acknowledgments}
This work is part of the FLASH project of PEPR LUMA and was supported by the French National Research Agency, as a part of the France 2030 program, under grant ANR-24-EXLU-0004. It was also funded by a public ANR grant referenced as EUR MANUTECH SLEIGHT-ANR-17-EURE-0026.  The numerical calculations were performed using computer resources from GENCI, project gen7041.

\end{acknowledgments}

\bibliography{apssamp}

\onecolumngrid

\setcounter{figure}{0}
\setcounter{table}{0}
\setcounter{equation}{0}
\renewcommand{\thefigure}{S\arabic{figure}}
\renewcommand{\thetable}{S\arabic{table}}
\renewcommand{\theequation}{S\arabic{equation}}

\vspace{70pt}
\section*{Supplemental Material}

\begin{figure}[H]
\centering
\includegraphics[scale=0.5]{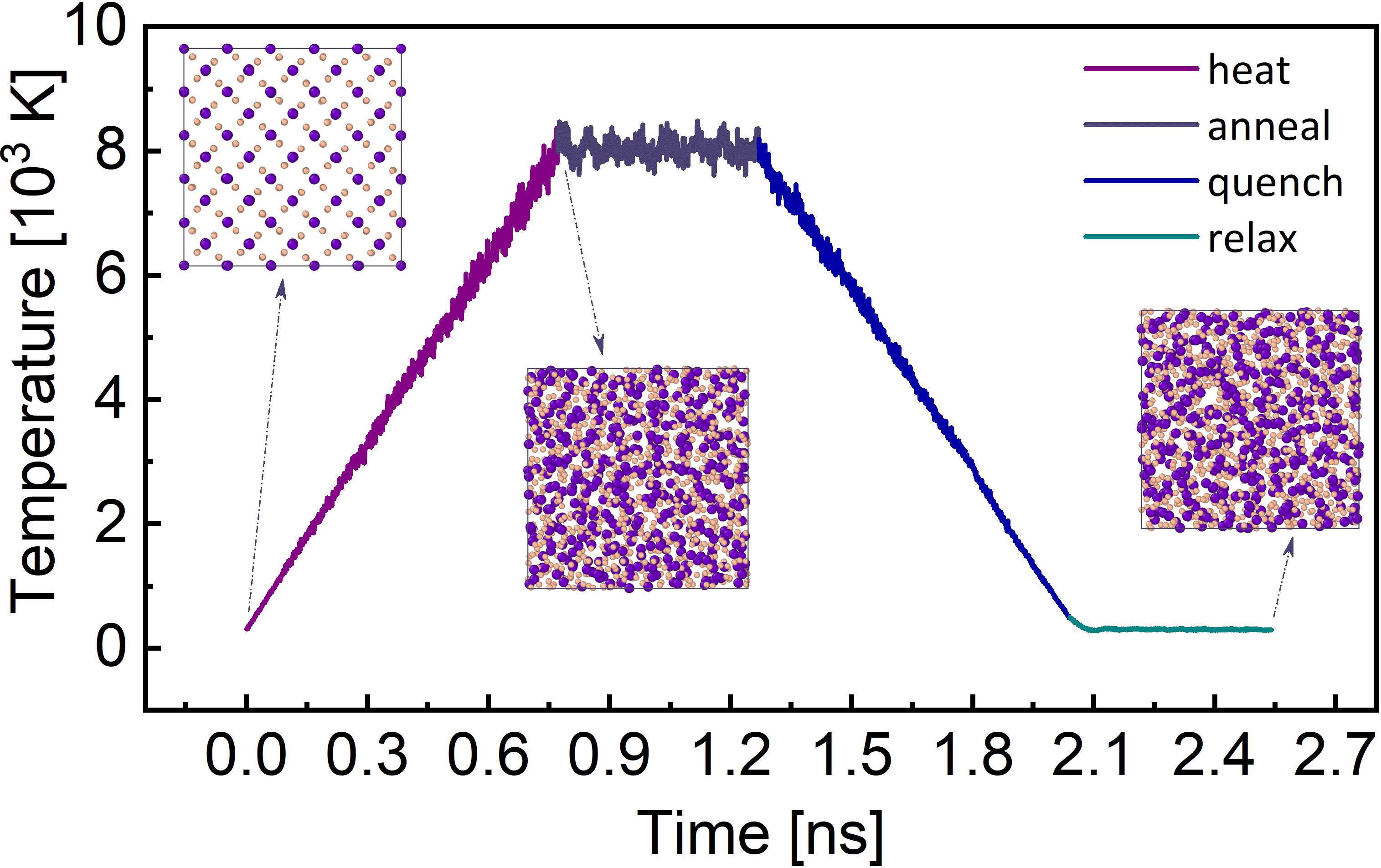}
\caption{Temporal evolution of temperature during the amorphization of stishovite. A standard heat-anneal-quench-relax protocol is used with canonical ensemble. The annealing stage is applied after heating to ensure a total removal of crystalline memory from the already amorphized system. A relaxation stage is applied to remove any residual stress after the quench. The snapshots show the initial ordered (c-axis) and the disordered configurations during the amorphization process.}
\label{T_t}
\end{figure}

\newpage

\begin{figure}[H]
\centering
\includegraphics[scale=0.61]{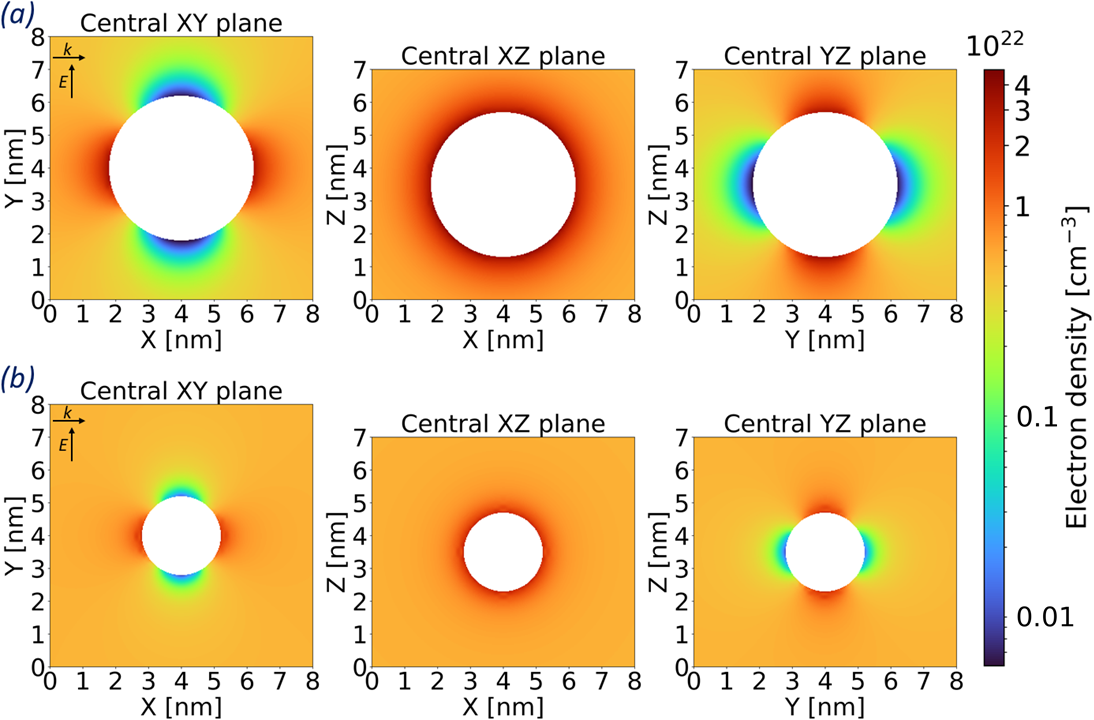}
\caption{Central slices of free-electron density distribution after laser irradiation, smoothened for better visualization, for the structure with a (a) 4-nm-pore and (b) 2-nm-pore.
The free-electron density is strongly modulated by the presence of the pore for both pore sizes, reaching a maximum value right at the pore surface boundary of approximately $4 \times 10^{22}$ cm$^{-3}$. The highest electron densities are localized at the pore surface regions aligned with the laser polarization direction. The discontinuity in permittivity at the dielectric-pore interface leads to local electric field amplification, which enhances nonlinear ionization processes and results in higher electron densities near the pore boundary.}
\label{free_electron_density}
\end{figure}

\begin{figure}[H]
\centering
\includegraphics[scale=0.5]{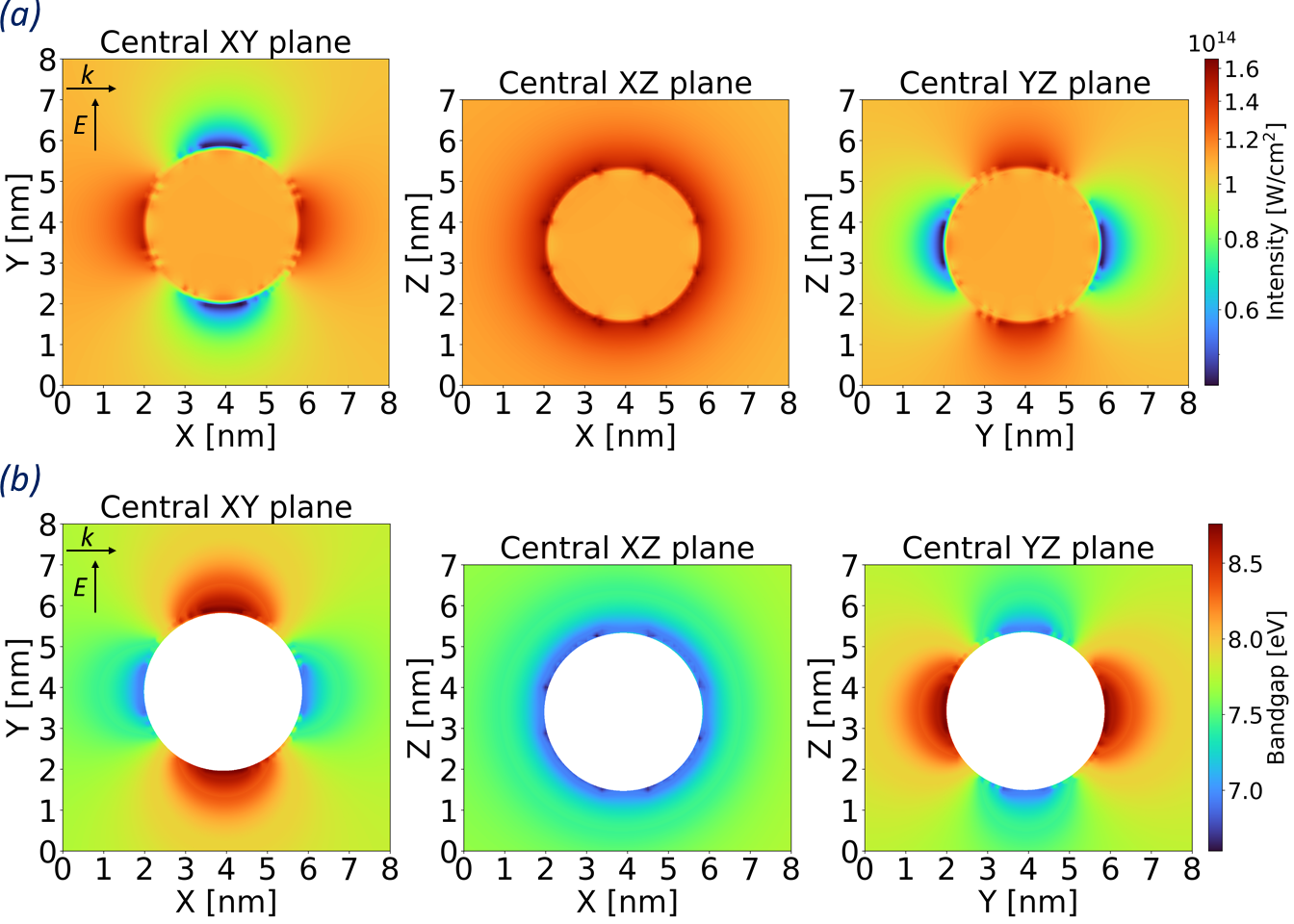}
\caption{(a) Smoothened intensity spatial distribution of 4-nm-pore structure taken from FDTD data when the laser pulse intensity was at its peak. (b) The corresponding bandgap distribution obtained by mapping the local intensity to the intensity-dependent bandgap (top panel in Fig. 2 of main text). The maximum local intensity exceeds the incident peak value by a factor of around 1.65 in the hottest spots of the pore boundaries, indicating significant near-field enhancement at silica interface. In the background, the bandgap decreases from the ground-state 8.9 eV to around 7.9 eV, while at the hottest spots along the pore boundary it further decreases to approximately 6.6 eV. This localized bandgap shrinkage is confined to a narrow region around the interface, corresponding to the steep spatial gradients of the electromagnetic field. Even at intensities close to approximately $4.5 \times 10^{13}$~W/cm$^{2} - 1.65 \times 10^{14}$~W/cm$^{2}$ range, relevant to our simulations, the dynamic bandgap increases the calculated rate at the hottest spots of the pore boundaries by a factor of up to about~6 compared to the static-bandgap case, indicating that field-driven gap renormalization significantly amplifies free-carrier generation. One should also note that while the ratio of ionization rate between the pore boundary and background homogeneous region was approximately 4 in case of static bandgap, the same ratio increased to approximately 10 in case of dynamic bandgap.}
\label{intensity_bandgap_colormap}
\end{figure}

\begin{figure}[H]
\centering
\includegraphics[scale=0.5]{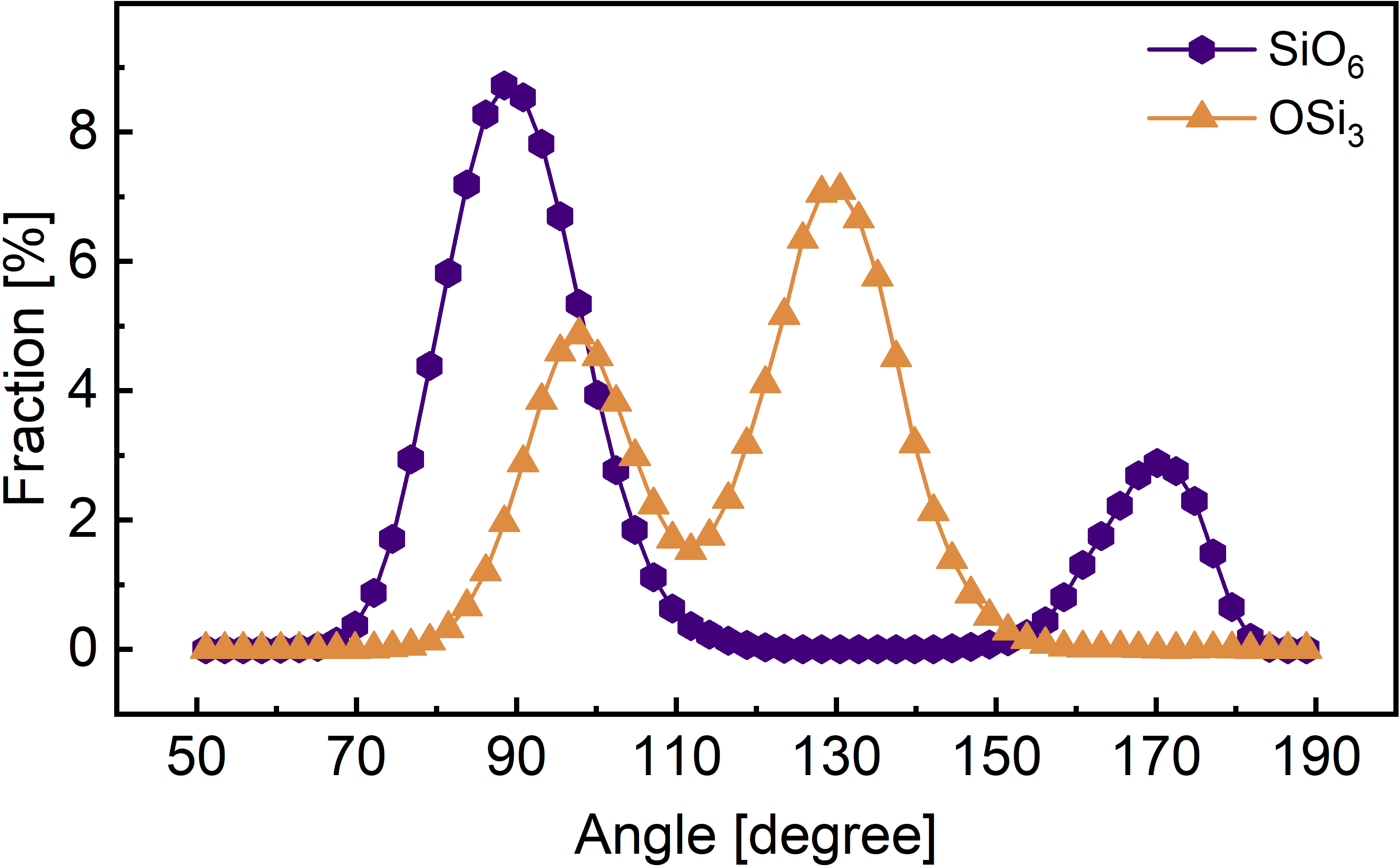}
\caption{Bond-angle distribution for O-Si-O and Si-O-Si angles within SiO$_6$ and OSi$_3$ coordination units of the crystallized system 2 ns after irradiation. The peaks at 90\textdegree $ $ and 170\textdegree $ $ of SiO$_6$ unit correspond to adjacent and trans ligands, respectively. Similarly, the peaks at 97\textdegree $ $ and 130\textdegree $ $ of OSi$_3$ unit show the edge sharing and corner-sharing connections.}
\label{SiO6_and_OSi3}
\end{figure}

\vspace{50pt}
\begin{figure}[H]
\centering
\includegraphics[scale=0.5]{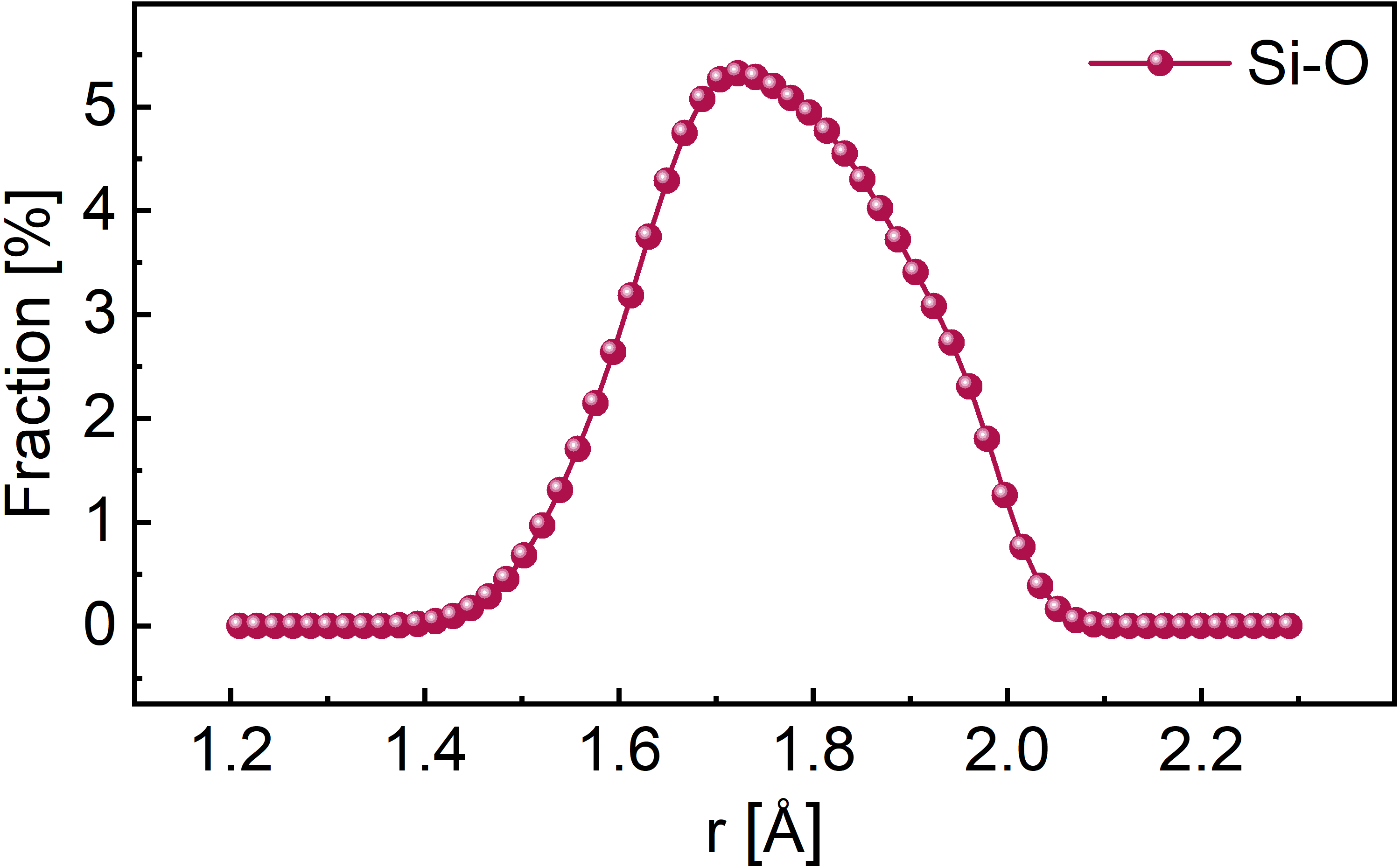}
\caption{Si-O bond-length distribution, peaked around 1.73 \AA, within SiO$_6$ coordination units in the crystallized system at a final pressure of 27 GPa.}
\label{SiO_bond_length}
\end{figure}

% \appendix

\end{document}